\documentclass{sig-alternate-10pt}
\usepackage[margin=1in]{geometry}

\usepackage{bm} 
\usepackage{url}
\usepackage{cite}
\usepackage{amssymb}
\usepackage{amsmath}
\usepackage{amsfonts}
\usepackage{algorithmic}
\usepackage{graphicx}
\usepackage{textcomp}
\usepackage{xcolor}
\usepackage{soul} 

\def\BibTeX{{\rm B\kern-.05em{\sc i\kern-.025em b}\kern-.08em
    T\kern-.1667em\lower.7ex\hbox{E}\kern-.125emX}}

\usepackage{hyperref}
\usepackage[colorinlistoftodos]{todonotes}

\usepackage[ruled, noend]{algorithm2e}
\usepackage[shortlabels]{enumitem}

\usepackage{tikz}

\usepackage{microtype}

\usepackage{relsize}

\newcommand{\nop}[1]{}

\newcommand{\set}[1]{\{#1\}}                    
\newcommand{\setof}[2]{\{{#1}\mid{#2}\}}        
\newcommand{\pr}{\mathop{\textnormal{Pr}}}    
\newcommand{\dom}{\textsf{Dom}}

\newcommand{\degree}{\texttt{deg}}

\newcommand{\dan}[1]{\todo[inline,color=yellow]{\textsf{#1} \hfill \textsc{--Dan.}}}

\newtheorem{thm}{Theorem}[section]
\newtheorem{lmm}[thm]{Lemma}

\newcommand{\defeq}{\stackrel{\text{def}}{=}}

\newcommand{\Rp}{{\mathbb R}_{\tiny +}} 

\newcommand{\ce}{\textsc{CE}\xspace}
\newcommand{\pce}{\textsc{PCE}\xspace}
\newcommand{\est}{\textsc{Est}}

\newcommand{\dsb}{\textsc{DSB}\xspace}
\newcommand{\cb}{\textsc{CB}\xspace}
\newcommand{\bs}{\textsc{BoundSketch}\xspace}
\newcommand{\agm}{\textsc{AGM}\xspace}
\newcommand{\polyb}{\textsc{PolyB}\xspace}
\newcommand{\attrs}{\texttt{Attrs}}

\newcommand{\lp}[1]{||#1||}

\begin{document}
\pagestyle{empty}

\title{Pessimistic Cardinality Estimation\thanks{This work was
    partially supported by NSF-BSF 2109922, NSF IIS 2314527, NSF SHF
    2312195, SNSF 200021-231956, and RelationalAI.}}

\numberofauthors{4} 
\author{
\alignauthor
Mahmoud Abo Khamis\\
       \affaddr{RelationalAI, Inc}
\alignauthor
Kyle Deeds\\
       \affaddr{\mbox{University of Washington}}
\and
\alignauthor
Dan Olteanu\\
       \affaddr{University of Zurich}
\alignauthor
Dan Suciu\\
       \affaddr{\mbox{University of Washington}}
}

\maketitle

\begin{abstract}
  Cardinality Estimation is to estimate the size of the output of
  a query without computing it, by using only statistics on the input
  relations.  Existing estimators try to return an unbiased estimate
  of the cardinality: this is notoriously difficult.  A new class of
  estimators have been proposed recently, called \emph{pessimistic}
  estimators, which compute a guaranteed upper bound on the query
  output.  Two recent advances have made pessimistic estimators
  practical.  The first is the recent observation that degree
  sequences of the input relations can be used to compute query upper
  bounds.  The second is a long line of theoretical results that have
  developed the use of information theoretic inequalities for query
  upper bounds.  This paper is a short overview of pessimistic
  cardinality estimators, contrasting them with traditional
  estimators.
\end{abstract}

\section{Introduction}

\bigskip

Given a query, the \emph{Cardinality Estimation} problem, or \ce for
short, is to estimate the size of the query output, using precomputed
statistics on the input database. This estimate is the primary metric
guiding cost-based query optimization. The optimizer uses it to make
decisions about every aspect of query execution. These range from
broad logical optimizations like the join order and the number of
servers to distribute the data over, to detailed physical
optimizations like the use of bitmap filters and memory allocation for
hash tables.

Unfortunately, \ce is notoriously difficult.  Density-based
estimators, pioneered by
System~R~\cite{DBLP:conf/sigmod/SelingerACLP79}, are widely used
today, but they underestimate significantly due to their strong
reliance on assumptions like data uniformity and independence.
They tend to have large errors for queries with many joins and many
predicates~\cite{DBLP:journals/pacmmod/DeedsSB23,DBLP:journals/pvldb/HanWWZYTZCQPQZL21}.
A major limitation of density-based \ce is that it does not come with
any theoretical guarantees\nop{ about its estimate}: it may under-, or
over-estimate, by a little or by a lot, without any warning.
Alternative approaches have been proposed, but they come with their
own limitations: sampling-based estimators require expensive data access at estimation time, while ML-based estimators \nop{tend to be
  complex,} suffer from large memory footprints and long training
time~\cite{DBLP:journals/pvldb/WangQWWZ21,DBLP:conf/sigmod/KimJSHCC22,DBLP:journals/pacmmod/DeedsSB23}.


An alternative to \emph{estimation} is to compute an \emph{upper
  bound} on the size of the query output, based on the precomputed
statistics on the data.  This is called \emph{pessimistic cardinality
  estimation}, or \pce for short~\cite{DBLP:conf/sigmod/CaiBS19}.  The
\pce returns a number that is larger than the size of the query output
for \emph{any} database instance that satisfies the given statistics.
The inclusion of the term ``estimation'' in \pce is a little
misleading, since it does not estimate but gives an upper bound, yet the
term has stuck in the literature, and we will use it too.  Of course,
we want the upper bound to be as small as possible, but not smaller.
When it achieves accuracy that is similar to, or even better than
traditional estimators, \pce offers several advantages.  Its
one-sided theoretical guarantee can be of use in many applications,
for example it can guarantee that a query does not run out of memory,
or it can put an upper bound on the number of servers required to
distribute the output data.  It also has the advantage that the upper bounds combine naturally: if we have multiple upper bounds due to different techniques, we can apply them all and take the minimum.

This paper gives a high level overview of pessimistic cardinality
estimation, and contrasts it with traditional methods.  We start by
discussing single-block SQL queries:
\begin{align}
& \texttt{SELECT *}\nonumber\\
& \texttt{FROM R1, R2, ...} \label{eq:sql:q} \\
& \texttt{WHERE [joins and filters]} \nonumber
\end{align}
Later we will restrict our discussion to conjunctive queries.  Some of
the \pce techniques described here also apply to $\texttt{group-by}$
queries.

\section{Brief Review of CE}

\label{sec:review}

\bigskip

Modern database engines use a {\em density-based} approach for
cardinality estimation.  They compute periodically some simple
statistics (or summaries) of the base relations, then use simplifying
assumptions to estimate the cardinality.  Specifically, the engine
stores in its catalog, for each relation $R$, its cardinality $|R|$,
and the number of distinct values\footnote{Denoted $V(R,X)$ in a
  popular textbook~\cite{DBLP:books/daglib/0020812}.} $|\dom(R.X)|$,
where $X$ is a single attribute or a set of attributes.  This quantity
is computed periodically, and approximately, by sampling from $R$
(e.g. Postgres) or by using Hyper-Log-Log (e.g. DuckDB).  The ratio
$\frac{|R|}{|\dom(R.X)|}$ represents the \emph{average degree}, or the
\emph{density} of the attribute(s) $X$.

Consider the uniform probability space whose outcomes are the tuples
of the Cartesian product of all relations in~\eqref{eq:sql:q},
$R_1 \times R_2 \times \ldots$ This defines a probability
$\pr(X, Y, \ldots)$ over all attributes returned by the query.  A
\emph{density-based} \ce estimates the probability that a random tuple
from the Cartesian product satisfies the condition in the
\texttt{where} clause.  The cardinality estimate is the multiplication of this probability with the size of the Cartesian product.
For example, consider the following SQL query $Q$:
\begin{align*}
  &\texttt{SELECT }*\texttt{ FROM Store }
    \texttt{WHERE City}=\texttt{'Seattle'}
\end{align*}
By the {\em uniformity assumption} and the \emph{containment of values assumption} the probability that a random
tuple satisfies the predicate is
$\frac{1}{|\dom(\texttt{Store}.\texttt{City})|}$.  
The cardinality estimate is $\est(Q) =
\frac{|\texttt{Store}|}{|\dom(\texttt{Store}.\texttt{City})|}$.

\emph{1-Dimensional Histograms} relax the uniformity assumption, by
storing separate statistics for each bucket of a histogram.
The default number of buckets is small (200 for
SQLServer~\cite{sql-server-histogram-buckets}, 100 for Postgres), and
is strictly limited (typically to
$1000-10000$~\cite{DBLP:journals/pvldb/WangQWWZ21}).

Joins are estimated using similar assumptions. For example, consider
the query $J$ given by:
\begin{align}
  &\texttt{SELECT * FROM R, S WHERE R.X}=\texttt{S.Y} \label{eq:q:join}
\end{align}
One estimate could be $|R|\cdot\frac{|S|}{|\dom(S.Y)|}$, because each
tuple in $R$ matches an estimated $\frac{|S|}{|\dom(S.Y)|}$ tuples in
$S$.  Or, symmetrically, $\frac{|R|}{|\dom(R.X)|}\cdot|S|$.
Density-based \ce returns their minimum\footnote{Justified by the {\em
    containment-of-values} assumption.}, usually written as:
\begin{align}
  \est(J) = & \frac{|R|\cdot |S|}{\max(|\dom(R.X)|,|\dom(S.Y)|)} \label{eq:est:join}
\end{align}

Finally, the estimate for a conjunction of predicates is computed by assuming independence between the predicates.
For example, the estimate of the following query $Q$
\begin{align*}
  &\texttt{SELECT * FROM Store} \\
  &\texttt{WHERE City}=\texttt{'Seattle' AND Zip}=\texttt{'98195'}
\end{align*}
is
$\est(Q) =
\frac{|\texttt{Store}|}{|\dom(\texttt{Store}.\texttt{City})|\cdot
  |\dom(\texttt{Store}.\texttt{Zip})|}$, which is an underestimate,
since that entire zip code is in Seattle.  
\emph{2-Dimensional Histograms} can capture correlations between attributes, but few systems support them.\footnote{There are several reasons why 2-d histograms are rarely
  used. (1) There are too many candidates: $n(n-1)/2$ possible histograms
  for a relation with $n$ attributes. The number of buckets along
  each dimension is limited to $\sqrt{1000}-\sqrt{10000}$. (2) It is
  unclear how to combine multiple 2-d
  histograms~\cite{DBLP:journals/vldb/MarklHKMST07}, e.g. in order to
  estimate a predicate on 3 attributes $\pr(X,Y,Z)$ from 2-d
  histograms on $XY, XZ, YZ$.}

A landmark paper~\cite{DBLP:journals/pvldb/LeisGMBK015} evaluated the cardinality estimators deployed in modern
database systems and their impact on the query optimizer.  It found that their \ce almost
always underestimates (because of the independence assumption) with
typical errors of up to $10^4$ for queries with many joins.  Later studies have confirmed these findings across a wide variety of workloads and database systems~\cite{DBLP:journals/pvldb/LeeDNC23, DBLP:conf/sigmod/ParkKBKHH20,DBLP:journals/pvldb/WangQWWZ21,DBLP:journals/pvldb/HanWWZYTZCQPQZL21,DBLP:conf/sigmod/KimJSHCC22}.


Given the importance of the problem and the limitations of density-based methods, many recent proposals have been published exploring alternatives to traditional density-based \ce. Two alternatives have attracted  particular interest: {\em sampling-based} \ce and {\em learned} \ce.

Sampling-based \ce methods compute an unbiased estimate without
requiring any assumptions.  \emph{Offline sampling} pre-computes a
uniform sample $R_{\text{sample}} \subseteq R$ of each relation, then
estimates the size of the query output over the base relations from
the size of the query output over the sample using the
Horvitz-Thompson's formula.\footnote{The size estimate is the
  multiplication of the size of the query output over the sample with the
  ratio $|R|/|R_{\text{sample}}|$ for each relation in the query.
  More robust estimates, such as
  \emph{bottom-k}~\cite{DBLP:conf/pods/Cohen23}, are not commonly used
  for \ce~\cite{DBLP:conf/sigmod/ChenY17}.}  This method can be very
accurate for queries over a single relation and easily supports
arbitrary user-defined predicates beyond equality and range
predicates.  However, it becomes ineffective for highly selective
predicates, because of \emph{sampling collapse}: when no sampled tuple
matches the query, then the system must return 0 or 1.  In particular,
this is a problem for joins, since their selectivity is relative to
the cartesian product of the input relations and therefore almost
always extremely low.\footnote{The estimate is still unbiased, over
  the random choices of the samples, but the standard deviation is
  high.}  \emph{Online sampling} addresses this issue by sampling only
tuples that join with already sampled tuples.  Based on this
principle, Wander
Join~\cite{DBLP:conf/sigmod/0001WYZ16,DBLP:journals/tods/LiWYZ19}
achieves remarkable accuracy~\cite{DBLP:conf/sigmod/ParkKBKHH20};
however, it requires access to an index on every join column, and has
a high latency.

Learned \ce aims to remove the assumptions of traditional \ce by
training an ML model that captures the complex correlations
empirically~\cite{DBLP:journals/pvldb/WangQWWZ21,DBLP:conf/sigmod/KimJSHCC22}.
{\em Data-driven} estimators compute a generative ML model for the probability $\pr(X,Y,\ldots)$ over all attributes returned by the query. This is trained on the database~\cite{DBLP:journals/corr/abs-2012-14743,DBLP:journals/pvldb/YangKLLDCS20}.
The model needs to represent {\em all} attributes, of {\em all}
relations. Intuitively, these models aim for a lossy compression of 
the full outer join of the database relations then estimate the
selectivity of the predicates in the query relative to it. This is an extremely ambitious approach, and it tends to require large models that either struggle with, or completely disallow queries that do not follow the schema's natural join structure, like self-joins.  {\em Query-driven estimators} compute a discriminative model for
$\est(Q)$.  The model is trained using a workload of queries and 
their true cardinalities. In general, ML-based estimates can be quite accurate on the training data, but they suffer from distribution drift, can be memory intensive (1MB to 1GB models are reported in the literature),
support only limited types of queries and predicates, and require full
retraining when the data changes, e.g. when a new relation is added~\cite{DBLP:journals/pvldb/HanWWZYTZCQPQZL21}.

\section{A Wish List}
\label{sec:wish:list}

\bigskip

Stepping back from existing estimators, we ask a more general
question: What properties do we wish a cardinality estimator to have?
We propose here six such properties, which we argue any good \ce
should have:

\begin{description}
\item[Accuracy/Speed/Memory:] It should have good accuracy, small
  estimation time, small memory footprint.

\item[Locality:] It should use statistics that are computed
  separately on each input relation.

\item[Composition:] It should be able to compute an estimate for a 
  query from the estimates for its subqueries; this is useful in
  bottom-up query optimizers.
  
\item[Combination:] It should be able to combine multiple sources of
  statistics on the database. Given two estimates $\est_1$ and
  $\est_2$ computed using different methods, or different statistics,
  one should be able to combine them to obtain a better estimate
  $\est$.

\item[Incremental Updates:] It should be possible to update the
  statistics incrementally when the input data is updated.

\item[Guarantees:] It should offer some theoretical guarantees.  This
  will allow the application to reason about decisions based on \ce.
\end{description}

All \ce's aim for good speed/accuracy/memory, with various degrees of
success.  Density-based and sampling-based \ce's are local, while
learned \ce are definitely not local: they are monolithic, in that
they require access to all relations at training time.  Density-based
\ce is compositional\footnote{Under the {\em preservation of values}
  assumption.}, but it cannot do combination.  For example, 
to estimate the size of $\sigma_{X=a \wedge Y=b \wedge Z=c}(R)$ we
could use $|\dom(R.X)|$ and $|\dom(R.YZ)|$, or we could use
$|\dom(R.XY)|$ and $|\dom(R.Z)|$, but we cannot combine these two
estimates to get a better
estimate~\cite{DBLP:journals/vldb/MarklHKMST07}.  For another example,
given the cardinalities of both joins $R \Join S$ and $S \Join T$,
there is no canonical way to combine them to estimate the cardinality
of $R \Join S \Join T$~\cite{DBLP:journals/pvldb/ChenHWSS22}.
ML-based estimators can be neither composed nor combined, and need to
be re-trained from scratch after an update.  Finally, of all methods
discussed here, only sampling-based methods offer theoretical
guarantees: the others offer no guarantees.

Next, we will discuss the alternative, pessimistic approach to
cardinality estimation.  We will return to our wish-list in
Sec.~\ref{sec:discussion}.

\section{Pessimistic CE}

\label{sec:pce}

\bigskip

A \emph{Pessimistic Cardinality Estimator}, \pce, computes a {\em
  guaranteed upper bound} on the cardinality, instead of an estimate.
For a very simple example, an upper bound of the
join~\eqref{eq:q:join} is $|R|\cdot |S|$.  If we know the largest
number $b$ of occurrences of any value in $S.Y$, also called the {\em
  maximum degree of $Y$ in $S$}, then a better bound is $|R| \cdot b$;
for example, if $Y$ is a key in $S$, then $b=1$ and the bound becomes
$|R|$.  Symmetrically, if we know the maximum degree of $Y$ in
$R$, call it $a$, then $a \cdot |S|$ is also an upper bound.  We can
combine these bounds by taking their $\min$:
\begin{align}
|J| \leq \min(|R|\cdot b,\ a\cdot |S|) \label{eq:naive:pce:join}
\end{align}
This should be compared with the traditional estimator~\eqref{eq:est:join}, which replaces the maximum degrees $a$ and $b$ with the average degrees.

Two advances make pessimistic cardinality estimators
practical today.  The first is the observation that degree sequences
can be used to compute an upper bound on the query
output size~\cite{DBLP:conf/icdt/DeedsSBC23}.  The second is a long line of
theoretical results on using information inequalities to compute upper
bounds on the query's
output~\cite{friedgut-kahn-1998,DBLP:conf/soda/GroheM06,DBLP:journals/talg/GroheM14,DBLP:journals/jacm/GottlobLVV12,DBLP:conf/pods/KhamisNS16,DBLP:conf/pods/Khamis0S17,DBLP:conf/lics/Suciu23,DBLP:journals/corr/abs-2306-14075}.
We review both topics in the rest of the paper.

\section{Degree Sequences}

\label{sec:degree:sequence}

\bigskip

The existing \pce's use statistics derived from
\emph{degree sequences} of the input relations.

Let $R$ be a relation, and $U,V \subseteq \attrs(R)$ two sets of
attributes.  We assume throughout the paper that relations are \emph{sets}.
The \emph{degree sequence} 
\begin{align}
  \degree_R(V|U) \defeq & (d_1, d_2, \ldots, d_n) \label{eq:ds}
\end{align}
is defined as follows.  If $u_1, \ldots, u_n$ are the distinct values
of $\Pi_U(R)$, then $d_i \defeq |\sigma_{U=u_i}(\Pi_{UV}(R))|$ is the
degree of the value $u_i$. We assume that the values
$u_i$ are sorted in decreasing order of their degrees, i.e.
$d_1 \geq d_2 \geq \cdots \geq d_n$.  We say that the value $u_i$ has
\emph{rank} $i$.  Notice that $\degree_R(V|U)=\degree_R(UV|U)$.

%
%

\begin{figure}
  \includegraphics[width=0.98\linewidth]{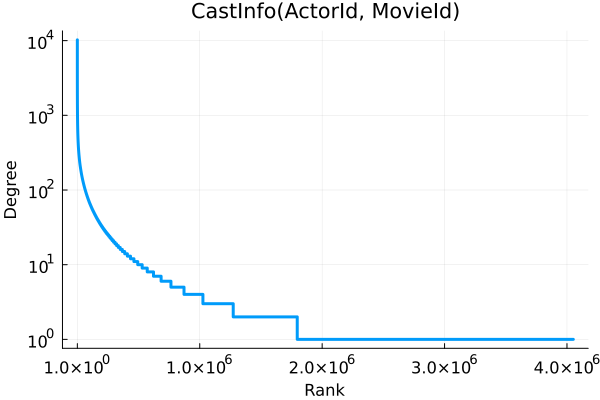}
\newline\null\hspace{5mm} $\lp{\degree_{\text{CastInfo}}(\text{MovieId}|\text{ActorID})}_1=36\cdot 10^6$
\newline\null\hspace{5mm}  $\lp{\degree_{\text{CastInfo}}(\text{MovieId}|\text{ActorID})}_\infty=10^4$
\caption{The degree sequence of $\text{CastInfo}$ from the JOB
  benchmark~\cite{DBLP:journals/pvldb/LeisGMBK015}.  Its cardinality
  is $36\cdot 10^6$, and it has $4\cdot 10^6$ distinct actor IDs, with
  degrees ranging from $10^4$ to $1$.  The maximum degree is that of
  Bob Barker, who hosted the CBS show \emph{The Price Is Right} from
  1972 to 2007 and also \emph{Truth or Consequences} from 1956 to
  1975.}
  \label{fig:degree:imdb}
\end{figure}

Figure~\ref{fig:degree} illustrates some simple examples of degree
sequences, while Figure~\ref{fig:degree:imdb} shows the degree
sequence of the IMDB dataset from the JOB
benchmark~\cite{DBLP:journals/pvldb/LeisGMBK015}.  If the functional
dependency $U \rightarrow V$ holds, then
$\degree_R(V|U)=(1,1,\ldots,1)$.  When $|U|\leq 1$ then we call the
degree sequence $\degree_R(V|U)$ {\em simple}, and when $UV=\attrs(R)$
are all the attributes of $R$, then we call $\degree_R(V|U)$ {\em
  full} and also denote it by $\degree_R(*|U)$.  The degree sequence
$\degree_R(YZ|X)$ in Fig.~\ref{fig:degree} is both simple and full,
and we can write it as $\degree_R(*|X)$.

\begin{figure}
  \centering
  {\scriptsize
    \begin{align*}
      R=&\begin{array}[c]{|c|c|c|} \hline X&Y&Z \\ \hline
         1&a&\ldots \\
         1&b&\ldots \\
         1&b&\ldots \\
         2&a&\ldots \\
         2&b&\ldots \\
         3&b&\ldots \\
         3&c&\ldots \\
         4&d&\ldots \\ \hline
    \end{array}
&&
   \begin{array}[c]{ll}
     \degree_R(YZ|X)&=(3,2,2,1)\\
     \degree_R(Y|X) &= (2,2,2,1)\\
     \degree_R(Z|XY)&=(2,1,1,1,1,1,1)\\
     \degree_R(XYZ|\emptyset)&=(8)=(|R|)
   \end{array}
    \end{align*}}
    \caption{Example of Degree Sequences}
  \label{fig:degree}
\end{figure}

The {\em $\ell_p$-norm} of the degree sequence~\eqref{eq:ds} is:
\begin{align*}
  \lp{\degree_R(V|U)}_p \defeq & \left(d_1^p + d_2^p + \cdots+d_n^p\right)^{1/p}
\end{align*}

Degree sequences and their $\ell_p$-norms generalize common statistics
used by density-based estimators: the relation cardinality is
$\lp{\degree_R(*|U)}_1=|R|$, the number of distinct values
$|\dom(R.U)|$ is the length $n$ of the sequence~\eqref{eq:ds} (and
also equal to $\lp{\degree_R(U|\emptyset)}_1$); and the maximum degree
of $R.U$ is $\lp{\degree_R(*|U)}_\infty=d_1$.  But degree sequences
contain information that can significantly improve the accuracy of
\pce.  For example the inequality $|J|\leq a \cdot |S|$
in~\eqref{eq:naive:pce:join} assumes pessimistically that all degrees
in $\degree_R(X|Y)$ are equal to the maximum degree $a$.  By using the
degree sequence and more sophisticated inequalities we can account for
the true distribution of the data, as we will see later.


In the rest of the paper we describe several methods that compute an
upper bound on the size of the query output from the $\ell_p$-norms of
degree sequences.  We restrict our discussion to conjunctive
queries instead of the SQL query~\eqref{eq:sql:q} and use the notation:
\begin{align}
  Q(X_1, \ldots, X_n) = & R_1(U_1) \wedge \ldots \wedge R_m(U_m)\label{eq:cq}
\end{align}
\noindent where $\set{X_1, \ldots, X_n} = U_1 \cup \cdots \cup
U_m$. The \emph{query variables} are $X_1, \ldots, X_n$. We omit
filter predicates for now, and discuss them in
Sec.~\ref{sec:pragmatic}.

\section{The AGM Bound}

\bigskip

Historically, the first cardinality upper bound was
introduced by Atserias, Grohe, and
Marx~\cite{DBLP:journals/siamcomp/AtseriasGM13}, and builds on earlier
work by Friedugut and Kahn~\cite{friedgut-kahn-1998} and Grohe and
Marx~\cite{DBLP:conf/soda/GroheM06}.  The bound is known today as the
\agm bound.  It uses only the cardinalities of the input relations,
$|R_1|, \ldots, |R_m|$, and is defined as follows.  A \emph{fractional
  edge cover} of the query $Q$ in~\eqref{eq:cq} is a tuple of
non-negative weights, $\bm w = (w_1, w_2, \ldots, w_m)$, such that
every variable $X_j$ is ``covered'', meaning
$\sum_{i: X_j\in U_i}w_i \geq 1$, for $1\leq j\leq n$.  Then, the following inequality holds:
\begin{align}
  |Q| \leq & |R_1|^{w_1}\cdot|R_2|^{w_2}\cdots |R_m|^{w_m} \label{eq:agm}
\end{align}
The \agm bound of $Q$ is defined as the minimum of~\eqref{eq:agm},
taken over all fractional edge covers $\bm w$.

The classical illustration of the \agm bound is for the 3-cycle query:
\begin{align}
  C_3(X,Y,Z) = & R(X,Y)\wedge S(Y,Z) \wedge T(Z,X) \label{eq:q:triangle}
\end{align}
One fractional edge cover is $(1,1,0)$, which proves
$|C_3|\leq |R|\cdot |S|$.  Other fractional edge covers are $(1,0,1)$,
$(0,1,1)$, and $(\frac{1}{2},\frac{1}{2},\frac{1}{2})$, and each leads
to a similar upper bound on $|C_3|$. The \agm bound is their
minimum:\footnote{These four fractional edge covers are called the
  \emph{vertices of the edge cover polytope}.  In~\eqref{eq:agm} it
  suffices to restrict the $\min$ to the vertices of the edge cover
  polytope, because any other fractional edge cover is $\geq$ than
  some convex combination of the vertices.}
{\scriptsize
\begin{align*}
  |C_3| \leq & \min\left(|R|\cdot |S|,\ |R|\cdot |T|,\ |S|\cdot |T|,\ \left(|R|\cdot|S|\cdot|T|\right)^{1/2}\right)
\end{align*}
}

The \agm bound enjoys two elegant properties: it is computable in
PTIME in the size of the query $Q$ (by solving a linear program), and
the bound is guaranteed to be tight.  The latter means that, for any
set of cardinalities $|R_1|, |R_2|, \ldots$ there exists a \emph{worst
  case database instance}, with the same cardinalities, where the size
of the query output is equal to the \agm bound, up to a rounding error
equal to a query-dependent multiplicative constant.\footnote{The
  constant is $\frac{1}{2^n}$, where $n$ is the number of variables.}
  
Despite these attractive theoretical properties, the \agm bound has not been adopted in practice, because it uses very limited statistics on the input data,
namely just the relation cardinalities.  In particular, for acyclic
queries the \agm bound is achieved by an integral (not fractional)
edge cover, and expression~\eqref{eq:agm} is a product of the
cardinalities of some relations.  For example, for the 2-way join query:
\begin{align}
  J_2(X,Y,Z) = & R(X,Y) \wedge S(Y,Z) \label{eq:j2}
\end{align}
The \agm bound is $|R|\cdot |S|$.  In contrast, a density-based
estimator uses both cardinalities and average degrees, as seen for
example in~\eqref{eq:est:join}, and for this reason it returns
a better estimate for $J_2$ than the \agm bound, even if it cannot
provide any guarantees.

\section{The Chain Bound}

\label{sec:cb}

\bigskip

The chain bound~\cite{DBLP:conf/pods/KhamisNS16}, \cb for short, uses
as statistics both cardinalities $|R_i|$, and maximum degrees
$\lp{\degree_{R_i}(V|U)}_\infty$.  It strictly generalizes the \agm
bound, but a popular simplification adopted by several implementations
is weaker than the \agm bound.  To describe \cb, it is convenient
to view each cardinality $|R_j|$ as a maximum degree, namely
$|R_j| = \lp{\degree_{R_j}(*|\emptyset)}_\infty$.  Thus, all
statistics are max degrees: $\lp{\degree_{R_{j_i}}(V_i|U_i)}_\infty$,
for $i = 1, k$.

Fix an ordering $\pi$ of the query variables.  We say that the pair
$(V|U)$ \emph{covers} the variable $X$ w.r.t. $\pi$, and write
$X \in_\pi (V|U)$, if $X \in V$ and all variables in $U$ strictly
precede $X$ in the order $\pi$.  For a simple example, if
$U=\emptyset$, then $(V|\emptyset)$ covers all variables in $V$; for
another example, if the order $\pi$ is $X, Y, Z$ then $(XZ|Y)$ covers
only $Z$.  If a set of non-negative weights $w_1, \ldots, w_k$ is a
\emph{fractional cover} of $(V_1|U_1), \ldots, (V_k|U_k)$, meaning
that it covers every variable $X$ (i.e.,
$\sum_{i: (X \in_\pi V_i|U_i)} w_i \geq 1$), then the following
holds (we give the proof in the Appendix):
\begin{align}
  |Q| \leq & \prod_{i=1,k}\lp{\degree_{R_{j_i}}(V_i|U_i)}_\infty^{w_i}\label{eq:cb}
\end{align}
\cb is the minimum of the quantity above over all fractional covers
$\bm w$ and variable orderings $\pi$, and is always an upper bound on
$|Q|$.  If $\pi$ is fixed, then \cb can be computed in PTIME by using
an LP solver, but minimizing over all $\pi$ requires exponential
time.\footnote{Computing \cb is NP-hard~\cite{hung-2024}.}  When all
statistics are cardinalities then \cb is independent on the choice of
$\pi$ and coincides with the \agm bound. When the statistics also
include max-degrees, then \cb is lower (better) than the \agm bound.

Several implementations have adopted a simplified version of \cb which
does not require the use of an LP solver, called \bs
in~\cite{DBLP:conf/sigmod/CaiBS19,DBLP:conf/cidr/HertzschuchHHL21} or
MOLP in~\cite{DBLP:journals/pvldb/ChenHWSS22}: we will use the term
\bs.  It corresponds to restricting the fractional cover $\bm w$
in~\eqref{eq:cb} to be an integral cover.\footnote{\bs also hash
  partitions the data in order to improve the
  estimate~\cite{DBLP:conf/sigmod/CaiBS19}.  We will use \bs to refer
  to integral \cb, without data partitioning.}  We describe \bs,
following the presentation in~\cite{DBLP:journals/pvldb/ChenHWSS22}.
Consider the following nondeterministic algorithm.
\begin{itemize}
\item Set $W := \emptyset$, $CB := 1$.
\item Repeatedly choose non-deterministically one statistics,
  $d_i \defeq \lp{\degree_{R_{j_i}}(V_i|U_i)}_\infty$, such that
  $U_i \subseteq W$.  Set $W := W \cup V_i$ and $CB := CB \cdot d_i$.
\item Stop when $W$ contains all query variables, and return $CB$.
\end{itemize}

\bs is the minimum value returned by all executions of the
non-deterministic algorithm.  This is equivalent to computing the
shortest path in the graph called $CEG_M$
in~\cite{DBLP:journals/pvldb/ChenHWSS22}: its nodes are all subsets
$W \subseteq\set{X_1,\ldots,X_n}$, and for every statistics
$d_i = \lp{\degree_{R_{j_i}}(V_i|U_i)}_\infty$ there is an edge with weight
$d_i$ from $W$ to $W \cup V_i$, for all $W \supseteq U_i$.  Then \bs
is the shortest\footnote{Weights along a path are multiplied.}  path
from $\emptyset$ to $\set{X_1, \ldots, X_n}$.  \bs still
requires exponential time.\footnote{When the statistics are restricted
  to cardinalities (no max-degrees), then \bs or, equivalently,
  MOLP~\cite{DBLP:journals/pvldb/ChenHWSS22}, is larger (worse) than
  the AGM bound, as it considers only integral covers.  The
  original definition of MOLP~\cite{DBLP:conf/icdt/JoglekarR16} (called MO in~\cite{DBLP:conf/icdt/JoglekarR16}) claims the opposite, a claim repeated
  in~\cite{DBLP:journals/pvldb/ChenHWSS22}.  However, that claim only
  holds under the assumption that one first regularizes the data, then
  computes a separate bound for each degree configuration, using both
  cardinalities and max-degree statistics.}

\bs is similar to a densi\-ty-based estimator, where the
average degrees are replaced by maximum degrees: this was noted
in~\cite{DBLP:journals/pvldb/ChenHWSS22}.  We illustrate this on an
example with a 3-way join query:
\begin{align}
  J_3(X,Y,Z,U) = & R(X,Y)\wedge S(Y,Z)\wedge T(Z,U) \label{eq:j3}
\end{align}
\bs is the minimum of the following quantities:\footnote{We omit
  non-optimal expressions, like $|R|\cdot|S|\cdot|T|$.}
\begin{align}
  |J_3| \leq & |R|\cdot \lp{\degree_S(Z|Y)}_\infty \cdot \lp{\degree_T(U|Z)}_\infty\label{eq:j3:cb}\\
  |J_3| \leq & \lp{\degree_R(X|Y)}_\infty \cdot |S|\cdot \lp{\degree_T(U|Z)}_\infty\nonumber\\
  |J_3| \leq &  \lp{\degree_R(X|Y)}_\infty\cdot\lp{\degree_S(Y|Z)}_\infty \cdot |T|\nonumber\\
  |J_3| \leq & |R| \cdot |T|\nonumber
\end{align}
Let us compare this with the density-based estimator described in 
Sec.~\ref{sec:review}, which, for $J_3$, is:
\begin{align*}
  \est(J_3) = & \frac{|R|\cdot|S|\cdot|T|}{\max(|R.Y|,|S.Y|)\cdot\max(|S.Z|,|T.Z|)}
\end{align*}
For illustration, assume that $|R.Y|\leq |S.Y|$ and $|S.Z|\leq |T.Z|$.
Then the estimate can be written as:
\begin{align*}
  \est(J_3) = & |R| \cdot \frac{|S|}{|S.Y|} \cdot \frac{|T|}{|T.Z|}
\end{align*}
This is the same expression as~\eqref{eq:j3:cb}, where the maximum
degrees are replaced with the average degrees.  


The main advantage of \cb and  its simplified variant \bs is their simplicity and resemblance to the density-based estimator.
A disadvantage is that their computation requires exponential time,
because we need to iterate over all variable orderings $\pi$.  Another
limitation is that, unlike the \agm bound, \cb is not tight: the
polymatroid bound described below can be strictly lower than \cb.  However, in the special case when the set of statistics is \emph{acyclic}, \cb is both tight and computable in PTIME.  In this case, there exist variable orderings such that, for every statistics
$\lp{\degree_{R_{j_i}}(V_i|U_i)}_\infty$, all variables in $U_i$ come before those in $V_i$, which implies $V_i\subseteq_\pi(V_i|U_i)$ for
$i=1,k$.  Such an ordering $\pi$ can be computed in linear time using topological sort. We then use an LP solver to compute the optimal fractional cover for  $\pi$.

\section{The Polymatroid Bound}

\label{sec:polyb}

\bigskip

The {\em Polymatroid Bound}~\cite{DBLP:conf/pods/Khamis0S17}, \polyb,
can use any $\ell_p$-norms of degree sequences to compute an upper
bound on the size of the query output.  The input statistics for
\polyb are norms $\lp{\degree_R(V|U)}_p$, for a relation $R$, sets of
attributes $U,V$ of $R$, and numbers\footnote{$p$ can be fractional.}
$p > 0$.  The system uses all available statistics to compute an upper
bound on the query's output.  \polyb strictly generalizes both \cb and
the \agm bound.  Its theoretical foundation lies in information
inequalities, which we will review shortly, after presenting some
examples of \polyb.

{\bf Examples} To warm up, consider again the two-way join
$J_2(X,Y,Z) = R(X,Y)\wedge S(Y,Z)$.  The following upper bounds on the size of the query output hold:
\begin{align}
  |J_2| \leq & |R|\cdot |S|\label{eq:pce:j:1} \\
  |J_2|\leq & |R|\cdot\lp{\degree_S(Z|Y)}_\infty\label{eq:pce:j:2} \\
  |J_2|\leq & \lp{\degree_R(X|Y)}_\infty \cdot |S| \label{eq:pce:j:2b}\\
  |J_2|\leq & \lp{\degree_R(X|Y)}_2\cdot\lp{\degree_S(Z|Y)}_2\label{eq:pce:j:3}
\end{align}
The \polyb is the minimum of these four quantities\footnote{If all
  statistics consists of $\ell_p$-norms with $p$ an integer, or
  $p=\infty$, then the best bound on $J_2$ is the minimum of
  inequalities~\eqref{eq:pce:j:2}-\eqref{eq:pce:j:3}.}.  The first
three inequalities are trivial (and were discussed in
Sec.~\ref{sec:pce}), while~\eqref{eq:pce:j:3} follows from
Cauchy-Schwartz.

\nop{
\dan{We know the complete story, but I don't know how much we want to
  say here.  If the statistics are restricted to $\degree_R(X|Y)$ and
  $\degree_S(Z|Y)$ (which include cardinalities), then we know exactly
  the set of all non-dominated inequalities for $J_2$, namely they are
  all H\"older inequalities
  $|J_2|\leq \lp{\degree(X|Y)}_p\lp{\degree(Z|Y)}_q$, where
  $1/p+1/q=1$.  Actually, the set of {\em all} inequalities are
  H\"older with $1/p+1/q\geq 1$, but those where $1/p+1/q>1$ are
  dominated by increasing either $p$ or $q$.  For
  example~\eqref{eq:pce:j:1} has $p=q=1$ and is dominated, e.g.  by
  $p=\infty, q=1$; this follows from the fact that $r \geq p$ implies
  $\lp{\degree_R(X|Y)}_r \leq \lp{\degree_R(X|Y)}_p$.  If we consider
  only non-dominated inequalities, then the only integral solutions
  for $(p,q)$ are $(1,\infty)$, $(\infty,1)$, and $(2,2)$.  So, the
  inequalities that we list above are all inequalities with integral
  norms.  We also know (from the PODS paper, in the Appendix) that
  there are instances $R, S$ where a non-integral H\"older is
  asymptotically better than all the integral ones.  So, if we state
  that these are all inequalities, we must mention the assumption that
  all statistics use integer norms.}  \dan{Concrete suggestion.  We
  could say something like this.  ``One can check that, if all
  statistics consists of $\ell_p$-norms with $p$ an integer, or
  $p=\infty$, then the best bound on $J_2$ is the minimum of
  inequalities~\eqref{eq:pce:j:2}-\eqref{eq:pce:j:3}''}
}

Consider now the 3-way join
$J_3(X,Y,Z,U)=R(X,Y)\wedge S(Y,Z) \wedge T(Z,U)$.  \polyb includes all
\cb inequalities (see Eq.~\eqref{eq:j3:cb}) and new inequalities using
$\ell_p$-norms, for example the following inequality holds for all
$p \geq 2$:
\begin{align}
  |J_3| \leq & \big(|R|^{p-2}\cdot \lp{\degree_R(X|Y)}_2^2 \nonumber\\
             &\cdot \lp{\degree_S(Z|Y)}_{p-1}^{p-1}\cdot\lp{\degree_T(U|Z)}_p^p\big)^{1/p} \label{eq:pce:j3}
\end{align}
For example, assume that the system has computed the
$\ell_1, \ell_2, \ell_3$ norms of all degree sequences, then it can
compute the expression above for both $p=2$ and $p=3$, and take their
minimum.

Considering the 3-cycle query in~\eqref{eq:q:triangle}, the following
inequalities hold:

{\small
\begin{align}
  |C_3| \leq & \left(|R|\cdot |S| \cdot |T|\right)^{1/2}\label{eq:pce:t:1} \\
  |C_3| \leq & \left(\lp{\degree_R(Y|X)}_2^2\cdot\lp{\degree_S(Z|Y)}_2^2\cdot\lp{\degree_T(X|Z)}_2^2\right)^{1/3}\label{eq:pce:t:2} \\
  |C_3| \leq & \left(\lp{\degree_R(Y|X)}_3^3\cdot\lp{\degree_S(Y|Z)}_3^3\cdot|T|^5\right)^{1/6}\label{eq:pce:t:3}
\end{align}
}

The first is the \agm bound; the others are new, and quite surprising.
This list is not exhaustive: one can derive many more inequalities,
using various $\ell_p$-norms.  \polyb is the minimum of all of them
(restricted to the available $\ell_p$-norms), and can
be computed using a linear program, as explained later.

Inequalities~\eqref{eq:pce:j3}-\eqref{eq:pce:t:3} do not appear to
have simple, elementary proofs.  Instead, they are proven using
Shannon inequalities.

{\bf Shannon Inequalities} We review briefly Shannon Inequalities,
which are special cases of entropic inequalities, or information
inequalities, and show how to use them to prove the inequalities
above.

Let $X$ be a finite random variable, with $N$ outcomes
$x_1, \ldots, x_N$, and probability function $\pr$.
Its \emph{entropy} is defined as:\footnote{Usually $\log$ is in base 2, but any base can be used without affecting any results.}
\begin{align*}
  h(X) \defeq & - \sum_i \pr(x_i) \log \pr(x_i)
\end{align*}
It holds that $0 \leq h(X) \leq \log N$, and $h(X)=\log N$ iff $\pr$
is uniform, i.e., $\pr(x_1)=\cdots=\pr(x_N)=1/N$.

Let $X_1,\ldots, X_n$ be $n$ finite, jointly distributed random
variables.  For every subset $U$ of variables, $h(U)$ denotes the
entropy of the joint random variables in $U$.  For example, we have
$h(X_1X_3)$, $h(X_2X_4X_5)$, etc.  This defines an \emph{entropic
  vector}, $\bm h$, with $2^n$ dimensions, one for each subset of
variables.  The following inequalities hold and are called
\emph{Shannon basic inequalities} (the second is called
\emph{monotonicity} and the third is called \emph{submodularity}):
\begin{align*}
  h(\emptyset) =&\ 0\\
  h(U\cup V) \geq &\  h(U)\\
  h(U)+h(V)\geq &\  h(U\cup V)+h(U\cap V)
\end{align*}
%

A vector $\bm h \in \Rp^{2^n}$ that satisfies Shannon basic
inequalities is called a {\em polymatroid}.  Every entropic vector is
a polymatroid, but the converse does not hold in
general~\cite{zhang1998characterization}.

We now have the tools needed to prove
inequalities~\eqref{eq:pce:j3}-\eqref{eq:pce:t:3}.  We illustrate
only~\eqref{eq:pce:t:1}, and defer the others to the Appendix.
Consider three input relations $R, S, T$, and let $C_3$ denote the
output to the query, i.e. $ C_3(x,y,z)$ iff $R(x,y)$, $S(y,z)$, and
$T(z,x)$.  Define the uniform probability space on $C_3$: there are
$|C_3|$ outcomes $(x,y,z)$, each with the same probability $1/|C_3|$.
Let $\bm h \in \Rp^8$ be its entropic vector.  Note that
$h(XYZ)= \log |C_3|$, because the probability distribution is uniform,
and $h(XY) \leq \log |R|$, because the support of the variables $XY$
is a subset of the relation $R$.  Then the following inequalities
hold, and imply~\eqref{eq:pce:t:1}:
\begin{align*}
  \log |R|& + \log |S| + \log |T| \geq \\
  \geq & \underline{h(XY) + h(YZ)}+ h(ZX) \\
  \geq & h(XYZ) + \underline{h(Y)+h(ZX)}\\
  \geq & 2h(XYZ) = 2 \log |C_3|
\end{align*}
We have applied Shannon submodularity inequality twice, and shown it
by underlying the used terms.

{\bf Entropic Constraints} To prove upper bounds that involve degree
sequences,
like~\eqref{eq:pce:j3},~\eqref{eq:pce:t:2},~\eqref{eq:pce:t:3}, we
need to relate their $\ell_p$-norms to the entropic vector.  

Let $X_1, \ldots, X_n$ be jointly distributed random variables, whose 
support is the finite relation $R(X_1, \ldots,$ $X_n)$. Let two sets of
variables $U,V$. Let $\bm h$ be
the entropic vector of the variables.  
The \emph{conditional entropy} of $U,V$ is defined as $h(V|U)\defeq h(UV)-h(U)$.  

The following was proven in~\cite{DBLP:journals/pacmmod/KhamisNOS24}, for
any $p >0$:
\begin{align}
  \frac{1}{p}h(U)+h(V|U) \leq & \log \lp{\degree_R(V|U)}_p \label{eq:h:p}
\end{align}
When $p=1$, the inequality becomes $h(UV)\leq \log |\Pi_{UV}(R)|$,
and when $p=\infty$, it becomes
$h(Y|X) \leq \log \lp{\degree_R(V|U)}_\infty$.  For a simple example,
using~\eqref{eq:h:p} we can prove~\eqref{eq:pce:j:3}:
\begin{align*}
  \log&\lp{\degree_R(X|Y)}_2 + \log \lp{\degree_S(Z|Y)}_2 \geq\\
  \geq& \left(\frac{1}{2}h(Y)+h(X|Y)\right)+\left(\frac{1}{2}h(Y)+h(Z|Y)\right)\\
  = & h(Y)+h(X|Y)+h(Z|Y)\\
  \geq & h(Y)+h(X|Y)+h(Z|XY) = h(XYZ) = \log|J_2|
\end{align*}
%
%
Inequalities~\eqref{eq:pce:j3}-\eqref{eq:pce:t:3} are proven
similarly: we include them in
Appendix~\ref{app:proofs:of:inequalities}.

{\bf Computing \polyb} So far, we introduced \polyb as the minimum
bound that can be obtained from Shannon inequalities and entropic
constraints~\eqref{eq:h:p}.  To compute it, we use an equivalent, dual
definition, as the maximum value of a linear program (LP).

Suppose $Q$ has $n$ variables $X_1, \ldots, X_n$.  The LP has $2^n$
real-valued variables, $h(U)$, representing the unknown polymatroid
vector $\bm h \in \Rp^{2^n}$.  The objective is to maximize
$h(X_1X_2\cdots X_n)$, given two sets of linear constraints:
\begin{itemize}
\item Statistics constraints: there is one inequality of
  type~\eqref{eq:h:p} for each available input statistics.
\item Shannon basic inequalities: the list of all submodularity and
  monotonicity inequalities.
\end{itemize}
\polyb is the optimal value of this linear program.  Intuitively, we maximize $\log |Q| = h(X_1\cdots X_n)$, while
constraining $\bm h$ to be a polymatroid that satisfies all
statistics~\eqref{eq:h:p}.
  
We illustrate with an example, by showing the linear program for the
query $C_3$ in~\eqref{eq:q:triangle}:

{\small
\begin{align*}
  \begin{array}{|rl|l}\cline{1-2}\cline{1-2} && \\
    \multicolumn{2}{|c|}{\texttt{maximize }  h(XYZ) \texttt{ subject to}} & \\
                                             && \\ \cline{1-2}\cline{1-2}
                                             && \\
    h(XY) \leq & \log |R| &\text{cardinality} \\
    h(XZ) \leq & \log |S| &\text{constraints} \\
    \ldots&&  \\
    \frac{1}{2}h(X)+h(Y|X) \leq & \log \lp{\degree_R(Y|X)}_2 &  \text{Other $\ell_p$}\\
    \frac{1}{3}h(X)+h(Y|X) \leq & \log \lp{\degree_R(Y|X)}_3 & \text{constraints}\\ 
    \ldots && \\ \cline{1-2}
                                             && \\
    h(X)+h(Y) \geq & h(XY) & \text{all Shannon} \\
    h(XY)+h(XZ) \geq & h(X)+h(XYZ) & \text{inequalities}\\
    \ldots&& \\ \cline{1-2}
  \end{array}
\end{align*}
}

\polyb has several attractive properties.  It strictly generalizes
both \cb and the \agm bound, which use only statistics given by
$\ell_1$ and $\ell_\infty$.  In addition, \polyb can also be used to
estimate output sizes for queries with group-by (and select distinct)
clauses; for this, we just need to adjust the objective of the linear
program to refer to the entropic term for the group-by attributes.
Access to more statistics is guaranteed to never make the bound worse; e.g., we can improve the bound if we decide to add the
statistics $\ell_3, \ell_4, \ell_5$ in addition to
$\ell_1, \ell_2, \ell_\infty$.
The inference time is exponential in $n$, the number of query
variables, but several special cases are known when it can be computed
in
PTIME~\cite{DBLP:conf/icdt/000122,DBLP:journals/corr/abs-2211-08381,LpBound-system}
(see Sec.~\ref{sec:pragmatic}).  Finally, when all statistics are
restricted to \emph{simple} degree sequences, then \polyb is provably
tight, up to a query-dependent multiplicative rounding
error,\footnote{The factor is $1/2^{2^n-1}$.} similar to the \agm
bound.  Beyond simple degree sequences, the \polyb is not tight in
general, due to the existence of non-Shannon
inequalities~\cite{zhang1998characterization,DBLP:conf/lics/Suciu23},
but the only known non-tight examples are artificial and unlikely to
occur in practice.

%
%

\section{The Degree Sequence Bound}

\label{sec:dsb}

\bigskip

The {\em Degree Sequence
  Bound}~\cite{DBLP:conf/icdt/DeedsSBC23,DBLP:journals/pacmmod/DeedsSB23},
\dsb for short, was the first system that proposed to use degree
sequences for \pce.  \dsb provides tighter bounds than both \agm and
\cb bounds, and an empirical
evaluation~\cite{DBLP:journals/pacmmod/DeedsSB23} found it to be often
tighter than density-based estimators, while always returning a
guaranteed upper bound.  Instead of using $\ell_p$-norms, \dsb uses
compressed representation of the degree sequences.  We illustrate here
\dsb assuming access to the full degree sequence, and discuss
compression in Sec.~\ref{sec:pragmatic}.


Consider the 2-way join $J_2(X,Y,Z) = R(X,Y) \wedge S(Y,Z)$, and assume that the
degree sequences of $R.Y$ and $S.Y$ are:
\begin{align*}
\degree_R(*|Y)=&(a_1, a_2, \ldots) & \degree_S(*|Y)=&(b_1,b_2, \ldots)
\end{align*}
Recall that the degree sequences are sorted, e.g.
$a_1 \geq a_2 \geq \cdots$, and that the rank of a value $y$ in $R.Y$
is the index $i$ for which the degree of $y$ is $a_i$.  If every value
$y$ has the same rank in $R.Y$ and in $S.Y$, then the size of the join
is precisely $|J_2| = \sum_i a_i b_i$.  In general, the following
inequality holds:
\begin{align}
  |J_2| \leq & \sum_i a_i b_i \label{eq:dsb:join}
\end{align}
%
This bound is a strict improvement over the chain
bound~\eqref{eq:naive:pce:join}, because
$\sum_i a_i b_i \leq a_1 \sum_i b_i = a_1 |S|$, and similarly
$\sum_i a_ib_i \leq |R| b_1$.

\dsb enjoys several nice properties.  It can be computed in linear
time in the size of the compressed degree sequences; it is
compositional, meaning that the estimate of a query plan can be
computed bottom-up; given appropriate histograms (see
Sec.~\ref{sec:pragmatic}) \dsb is more accurate than density-based
estimators~\cite{DBLP:journals/pacmmod/DeedsSB23}; finally, \dsb is
provably a tight upper bound of the query's output.  \dsb also has a
few limitations.  It is limited to Berge-acyclic queries
(see~\cite{DBLP:journals/jacm/Fagin83}), and its estimate is not given
in terms of an inequality, like \polyb, but instead it is computed by
an algorithm.

%
%

\section{Pragmatic Considerations}

\label{sec:pragmatic}

\bigskip

We discuss here some practical aspects that need to be considered by
\pce, or have been considered in previous implementations of
\pce~\cite{DBLP:conf/sigmod/CaiBS19,DBLP:conf/cidr/HertzschuchHHL21,DBLP:journals/pacmmod/DeedsSB23,LpBound-system}.

{\bf Statistics selection} While a density-based estimator only stores
the number of distinct values $|\dom(R.X)|$ of an attribute, \polyb
can use multiple $\ell_p$-norms of various degree sequences
$\degree_R(V|U)$.  A reasonable choice is to store, say, four numbers
$\ell_2, \ell_3, \ell_4, \ell_\infty$.  We assume that $|R|$ is stored
separately and do not include $\ell_1$.  This means that the memory
footprint increased by 4 times that of a density-based estimator.
But, in general, it is not clear whether we want to stop at
$\ell_4$. An implementation needs to choose which $\ell_p$ norms to
precompute.  Empirically, we have observed improvements of accuracy up
to $\ell_{30}$ for JOB queries on the IMDB dataset, although with diminishing returns~\cite{LpBound-system}.
On the other hand, \dsb compresses the degree sequence, and requires
tuning the hyperparameters of the compression.

{\bf Computing the statistics} All input statistics are computed
offline.  For example, the degree sequence $\degree_R(*|X)$ can be
computed using a group-by query:
\begin{align}
  & \texttt{SELECT } X, \texttt{COUNT(*)} \texttt{ FROM } R \texttt{ GROUP BY } X \label{eq:sql:ds}
\end{align}
\dsb sorts the query output, while \polyb computes all desired $\ell_p$-norms in one pass over this output.  
%

{\bf Conditional statistics} In order to improve the estimates of
queries with predicates, a pessimistic estimator can adopt traditional techniques like Most Common Values (MCV) and histograms.
Assume that the system decides to store the following \emph{global
  statistics} for column $X$: $\lp{\degree_R(*|X)}_p$ for
$p=2,3,4,\infty$, for a total of 4 numbers (recall that $\ell_1$ is
the cardinality which we store anyway).  To support equality
predicates $A=a$ on some attribute $A$, the system  computes
additional \emph{conditional statistics}:
\begin{itemize}
\item For each value $a \in \Pi_A(R)$, compute the degree sequence
  $\degree_{\sigma_{A=a}(R)}(*|X)$, which we denote by
  $\degree_R(*|X,A=a)$:
\begin{align}
  & \texttt{CREATE TABLE } DS \texttt{ AS }  \texttt{SELECT } A, X,  \label{eq:sql:conditional-ds} \\
  & \texttt{COUNT(*)} \texttt{ as } C \texttt{ FROM } R \texttt{ GROUP BY } A,X; \nonumber 
\end{align}
Then, compute the $\ell_p$-norms of these degree sequences (shown only for $\ell_2$ below):
\begin{align}
& \texttt{SELECT } A, (\texttt{SQRT}(\texttt{SUM}(C*C))) \texttt{ AS } L_2 \\
& \texttt{ FROM } DS \texttt{ GROUP BY } A; \nonumber 
\end{align}

\item For each Most Common Value (MCV) $a$, store its $\ell_p$-norms
  in the catalog.  Call these \emph{conditional MCV statistics}.  If
  we used $100$ MCVs (which is the default in Postgres), then these
  statistics consist of 400 numbers.
\item Compute the \emph{conditional common statistics},
  $\max_a \lp{\degree_R(*|X,A=a)}_p$, and add these 4 numbers to the
  catalog.
\item At estimation time, consider three cases.  If the query does not
  have a predicate on $A$, then use the global statistics; if the
  predicate is $A=a$, where $a$ is an MCV, then use the corresponding
  conditional MCV statistics; otherwise use the conditional common
  statistics.
\end{itemize}
%

{\bf Histograms} Histograms can further improve the accuracy.
Continuing the example above, we partition the values of the attribute
$A$ into buckets, and for each bucket store  conditional common
statistics: $\max_{a \in \texttt{Bucket}}\lp{\degree_R(*|X,A=a)}_p$.
There are 4 numbers per bucket. 
%
%
%
To support range predicates we need to store in each bucket the value
$\lp{\degree_R(*|X,A\in \texttt{Bucket})}_p$.
%

{\bf Boolean Expressions} The most powerful advantage of pessimistic
estimators is that the Boolean connectives $\wedge, \vee$ simply
becomes $\min$ and $+$.  For example, if the query contains predicate
$R.A = a \wedge R.B=b$, then we use as statistics
$\min(\lp{\degree_R(*|X,A=a)}_p,\lp{\degree_R(*|X,B=b)}_p)$;
similarly, for $R.A = a \vee R.B= b$, we use\footnote{
  We need to require $p \geq 1$, because Minkowski's inequality
  $\lp{\bm a + \bm b}_p \leq \lp{\bm a}_p + \lp{\bm b}_p$ only holds
  for $p \geq 1$.}  $+$.  This principle extends to \texttt{IN}
predicates (which are equivalent to $\bigvee$), and to \texttt{LIKE}
predicates, which can be upper-bounded by a set of 3-grams,
see~\cite{DBLP:journals/pacmmod/DeedsSB23}.

%

{\bf Lossy Compression} The complete degree sequence can be as large
as the data, and therefore it is not usable as a statistic for
cardinality estimation.  Since \dsb needs access to the entire degree
sequence, it uses three observations to drastically reduce the size of
the statistics.  First, \emph{run-length} compression is very
effective on degree sequences: it compresses a sequence
$\bm a = (a_1 \geq a_2 \geq \cdots)$ by replacing any constant
subsequence $a_i=a_{i+1}=a_{i+2}=\cdots$ with the common value plus
length.  Second, \dsb uses a lossy compression, replacing the original
sequence $\bm a$ with $\bm a'$ such that $\bm a \leq \bm a'$
(element-wise) and the compression of $\bm a'$ is much smaller: since
the \dsb is monotone in the input degree sequences, it still returns
an upper bound when $\bm a$ is replaced with $\bm a'$.  However, since
$\lp{\bm a}_1<\lp{\bm a'}_1$ , the new sequence represents a relation
with a larger cardinality.  Instead, \dsb upper bounds the CDF of the
degree sequence, $A_i \defeq \sum_{j \leq i} a_i$, instead of the PDF.
That is, it uses a degree sequence $\bm a''$ such that
$\bm A \leq \bm A''$ where $A_i'' = \sum_{j\leq i}a_i''$.  It is no
longer obvious why \dsb is still an upper bound when it uses
$\bm a''$: we give the main intuition in
Appendix~\ref{app:cdf:compression}.  Fig.~\ref{fig:compress}
illustrates this method.  The original sequence,
$\bm a = (4,2,2,1,1,1)$, can be compressed to $\bm a'=(4,4,4,2,2,2)$,
but the $\ell_1$ norm increased from 11 to 18.  Instead, \dsb upper
bounds the CDF to $\bm A \leq \bm A''$, see the second graph.  The
degree sequence that produces $\bm A''$ is
$\bm a'' = (4, 3.5, 3.5, 0, 0, 0)$: notice that
$\bm a \not\leq \bm a''$, yet by using $\bm a''$ \dsb still returns an
upper bound.

\begin{figure}
\begin{minipage}{0.1\linewidth}
{\scriptsize
$\begin{array}{|c|} \hline
      R.X \\ \hline
      a \\
      a \\
      a \\
      a \\ \hline
      b \\
      b \\ \hline
      c \\
      c \\ \hline
      d \\ \hline
      e \\ \hline
      f \\ \hline
    \end{array}
    $}
  \end{minipage}
  \begin{minipage}{0.9\linewidth}
    \includegraphics[width=\linewidth,keepaspectratio,trim={0 20mm 0 20mm},clip]{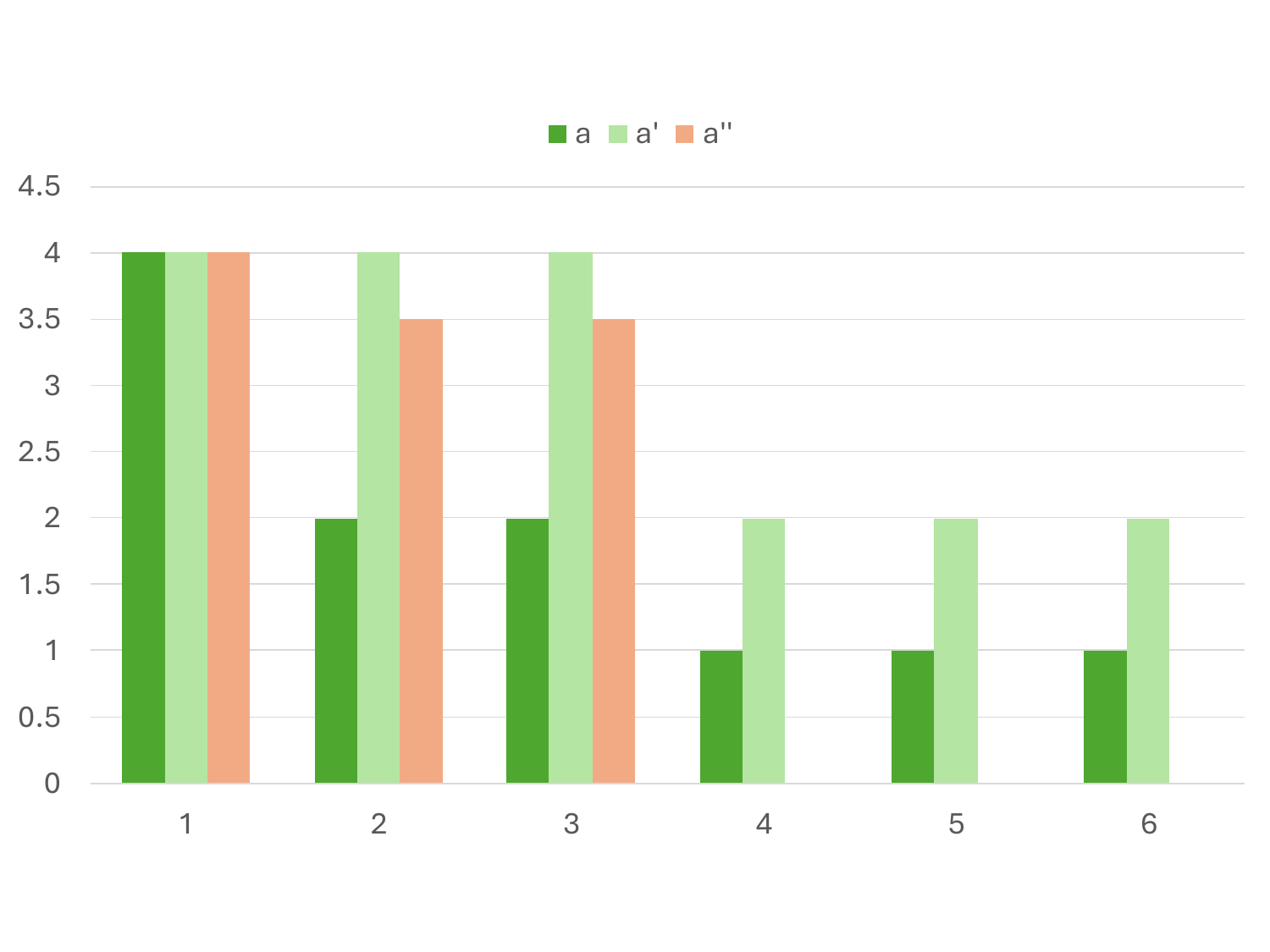}

    \includegraphics[width=\linewidth,keepaspectratio,trim={0 20mm 0 20mm},clip]{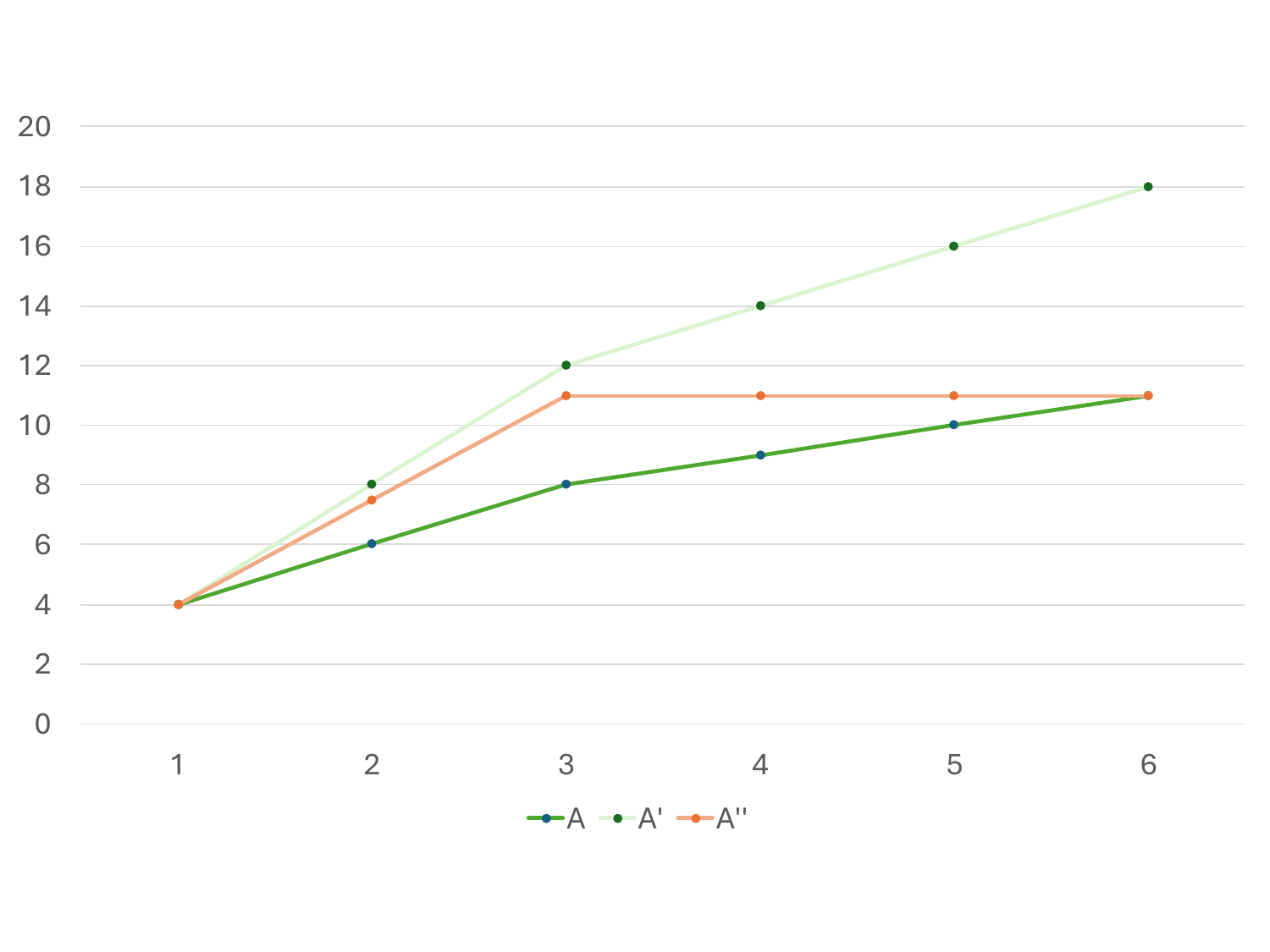}
  \end{minipage}
  \caption{Illustration of the advanced compression in \dsb: it shows
    the PDFs $\bm a$, $\bm a'$, $\bm a''$ and their CDFs
    $\bm A, \bm A', \bm A''$ (see text).}
  \label{fig:compress}
\end{figure}


{\bf Runtime of the Pessimistic Estimator} The runtime of \dsb is
linear in the size of the compressed representation of all degree
sequences.  The runtime of \polyb is exponential in $n$ (the number of
query variables, Eq.~\eqref{eq:cq}), because the LP uses $2^n$
numerical variables.  This is undesirable, but there are a few ways to
mitigate this.  First, when all statistics are for \emph{simple}
degree sequences (meaning of the form $\degree_R(V|X)$, where $X$ is a
single variable, or $X=\emptyset$), then \polyb can be computed in
PTIME, by using a totally different LP, with $n^2$ numerical variables
instead of $2^n$~\cite{DBLP:journals/corr/abs-2211-08381,LpBound-system}; furthermore, we only need $O(n)$ numerical variables in case of Berge acyclic queries~\cite{LpBound-system}.
Restricting the pessimistic estimator to simple degree sequences is
quite reasonable, because many current systems already restrict their
statistics to a single attribute, e.g. they store $|\dom(R.X)|$, but
rarely $|\dom(R.XY)|$.  When multi-attribute statistics are needed,
then there are a few ways to optimize the LP that computes \polyb.
First, ensure that it includes only the \emph{elemental Shannon
  inequalities}~\cite[Chap.14.1]{Yeung:2008:ITN:1457455}: there are
$n(n-1)2^{n-3}+n$ elemental inequalities.  Second, drop unnecessary
variables.  For example, to compute the \polyb of a 2-way join
$R(X,Y,Z,U,V)\Join S(V,W,K,L)$ it suffices to keep only one
non-joining variable in each relation, i.e. simplify the query to
$R(X,V)\Join S(V,W)$.  Third, it is well known that creating the input
for the LP solver often takes more time than the solver itself: to
speedup the creation of the LP, the system can precompute and
represent all elemental Shannon inequalities, since these depend only
on $n$ and not on the query.

\section{A Wish Come True?}

\label{sec:discussion}

\bigskip

We return now to our wish list in Sec.~\ref{sec:wish:list} and examine
how pessimistic cardinality estimators do or do not satisfy them.  The
main advantage of \pce's is their \emph{one-sided guarantee}.  An
application can count on the fact that the query  output size will never
exceed the \pce bound.  The ability to provide a strong guarantee
without any assumption is the main advantage of \pce over both
traditional methods and ML-based methods.  They are also \emph{local},
in the sense that they only use statistics collected on each relation
independently: if a new relation needs to be added to the database
schema, then we only need to add statistics for that relation. This is
similar to traditional \ce's, and different from estimators based on
trained models, which need access to the entire database.

The next question is how accurate the \pce's are.  Only a few
empirical evaluations have been done: for
\bs~\cite{DBLP:conf/sigmod/CaiBS19,DBLP:journals/pvldb/ChenHWSS22,DBLP:conf/sigmod/ParkKBKHH20},
\dsb~\cite{DBLP:journals/pacmmod/DeedsSB23}, and
\polyb~\cite{LpBound-system}.  Out of the box, \bs produces worse
estimates than traditional estimators, however all implementations
augment \bs with
hash-partitioning~\cite{DBLP:conf/sigmod/CaiBS19,DBLP:conf/sigmod/ParkKBKHH20},
which improves the accuracy, but at the cost of increased memory and
significantly increased construction and estimation time.\footnote{An
  improved partitioning is described recently
  in~\cite{DBLP:journals/corr/abs-2405-06767}.} While \bs uses only
cardinalities and max degrees, \dsb and \polyb use the complete degree
sequences (under lossy compression) and norms on these degree
sequences, respectively, and produce much better bounds with less
memory and with small estimation
time~\cite{DBLP:journals/pacmmod/DeedsSB23,LpBound-system}.  What
several implementations report is that, when the traditional estimator
in Postgres is replaced by \bs, \dsb, or \polyb, the performance of
expensive SQL improves, sometimes significantly, because it causes the
optimizer to be more cautions in choosing an execution plan.  Cheap
queries, however, tend to suffer from a regression.

In general, the empirical evaluation of \pce's is still very limited.
Future research is needed to
evaluate the impact of indices, on the essential query
subexpressions, or of bitmap filtering,
see~\cite{DBLP:journals/pvldb/LeeDNC23} for an extensive discussion of
the impact of the cardinality estimator on the query optimizer.

Moving down on our wish list, we note that \dsb is
\emph{compositional}; in fact the algorithm computing
\dsb~\cite{DBLP:conf/icdt/DeedsSBC23,DBLP:journals/pacmmod/DeedsSB23}
is defined by recursion on the structure of a Berge-acyclic query.  On
the other hand, \polyb is \emph{not} compositional.  This appears to
be unavoidable for cyclic queries, which do not admit a recursive
definition.  For example, computing the upper bounds on the join
$R \Join S$ does not help us in deriving the bound
$\left(|R|\cdot|S|\cdot|T|\right)^{1/2}$ of the 3-cycle $C_3$
(Eq.~\eqref{eq:pce:t:1}).  An open question is whether \polyb can be
made compositional on (Berge-) acyclic queries.

Pessimistic cardinality estimators can be \emph{combined} naturally.
If we have two bounds on the query, $B_1$ and $B_2$, we can
``combine'' them by taking their $\min$.  Similar combinations can be
done to handle complex predicates, consisting of \texttt{AND},
\texttt{OR}, \texttt{IN}, \texttt{LIKE}.  We believe that this is the
second main advantage of \pce's over traditional or ML-based methods.
None of the other cardinality estimation methods have a natural way of
combining estimates, and a recent study has found that different
combination heuristics lead to quite different
accuracies~\cite{DBLP:journals/pvldb/ChenHWSS22}.

Finally, we consider the desire for \emph{incremental updates}.
Ideally, when the database is updated (via an insert/update/delete
query), we would like to update the statistics incrementally.  For
context, we point out that, to the best of our knowledge, no
relational database system updates its statistics incrementally,
because it would slow down OLTP workloads, especially because updating
a statistics requires acquiring a lock, which can quickly lead to high
contention.  A similar reason would prevent them to do incremental
updates for pessimistic cardinality estimators.  However, the question
remains whether the data structures used by \pce's are able to support
incremental updates.  This appears to require materializing the degree
sequence computed by Query~\eqref{eq:sql:ds}, which is unlikely to be
practical.  Future work is needed to explore whether
$\ell_p$-sketches~\cite{DBLP:conf/stoc/AlonMS96,cormode2020small} can
be adapted for the purpose of incremental updates of the input
statistics.

{\bf Acknowledgments} We thank Nicolas Bruno, Surajit Chaudhuri, Bailu
Ding, Benny Kimelfeld, Kukjin Lee, and Hung Ngo for their comments
that helped improve the paper.


{\small
\bibliographystyle{acm}
\bibliography{bib}
}

\appendix

\section{Proof of the Chain Bound}

\bigskip

We use the notations introduced in Sec.~\ref{sec:cb}.  Let $\bm w$ be
a fractional cover, and let $X_n$ be the last variable in the order
$\pi$.  We prove by induction on $n$:
\begin{align}
  E \defeq \sum_{j=1,k} w_jh(V_j|U_j) \geq & h(X_1\cdots X_n) \label{eq:shannon:cb}
\end{align}
For $j \in J_1\defeq \setof{j}{X_n\in V_j}$, set
$V_j' \defeq V_j-\set{X_n}$; then
$h(V_j|U_j)=h(V_j'|U_j)+h(X_n|U_jV_j')$.  For $j\not\in J_1$, let
$V_j' \defeq V_j$.  Then $\bm w$ is a fractional cover of
$(V_j'|U_j)$, and by induction we have
$E \geq h(X_1\cdots X_{n-1}) + \sum_{j \in J_1} w_j h(X_n|X_1\cdots
X_{n-1}) \geq h(X_1\cdots X_n)$ because $\sum_{j \in J_1} w_j \geq 1$.
The chain bound~\eqref{eq:cb} follows from~\eqref{eq:shannon:cb} using
the standard argument from Sec.~\ref{sec:polyb}.

\section{Some Inequality Proofs}

\label{app:proofs:of:inequalities}

\bigskip

We prove~\eqref{eq:pce:j3}-\eqref{eq:pce:t:3}.
Inequality~\eqref{eq:pce:j3} follows from:
{\scriptsize
\begin{align*}
  \begin{array}{r}
    \multicolumn{1}{l}{(p-2)\log|R|+2\log\lp{\degree_R(X|Y)}_2}\\
    + (p-1)\log\lp{\degree_S(Z|Y)}_{p-1} \\
    + p\log\lp{\degree_T(U|Z)}_p\\ \\
  \multicolumn{1}{l}{\geq(p-2)h(XY)+2\left(\frac{1}{2}h(Y)+h(X|Y)\right)}\\
  + (p-1)\left(\frac{1}{p-1}h(Y)+h(Z|Y)\right)\\
  +p\left(\frac{1}{p}h(Z)+h(U|Z)\right)\\ \\
  \multicolumn{1}{l}{=(p-1)h(XY)+h(X|Y)}\\
  +h(YZ)+ (p-2)h(Z|Y)\\
  + h(UZ)+(p-1)h(U|Z)\\ \\
  \multicolumn{1}{l}{= (p-2)\left[h(XY)+h(Z|Y)+h(U|Z)\right]}\\
  +\left[h(XY)+h(UZ)\right]\\
  +\left[h(YZ)+h(X|Y)+h(U|Z)\right]\\ \\
  \multicolumn{1}{l}{\geq p\cdot h(XYZU)=p\log |J_3|}
  \end{array}
\end{align*}}

We have proven inequality~\eqref{eq:pce:t:1} in Sec.~\ref{sec:polyb}.
Inequalities~\eqref{eq:pce:t:2} and ~\eqref{eq:pce:t:3} follows from:
{\scriptsize
  \begin{align*}
  &(h(X)+2h(Y|X))+(h(Y)+2h(Z|Y))+ (h(Z)+2h(X|Z))\\
  =&\left(h(XY)+h(Y|X)\right)+\left(h(YZ)+h(Z|Y)\right)+\left(h(XZ)+h(X|Z)\right)\\
  \geq & 3h(XYZ)
\end{align*}
\begin{align*}
    \begin{array}{l}
    \left(h(X)+3h(Y|X)\right)+\left(h(Z)+3h(Y|Z)\right)+5h(XZ) \\
    =\left(h(XY)+2h(Y|X)\right)+\left(h(YZ)+2h(Y|Z)\right)+5h(XZ)\\
    =2\left(h(XZ)+h(Y|X)\right)+2\left(h(XZ)+h(Y|Z)\right)\\
    \multicolumn{1}{r}{+\left(h(XY)+h(YZ)+h(XZ)\right)}\\
    \geq 2h(XYZ) + 2h(XYZ) + 2h(XYZ) = 6h(XYZ)
    \end{array}
  \end{align*}
}

\section{CDF Compression}

\label{app:cdf:compression}

\bigskip

We prove here that the upper bound in Eq.~\eqref{eq:dsb:join}
continues to hold if we replace the sequences $\bm a, \bm b$ with two
sequences $\bm a', \bm b'$ whose CDFs are larger.

It will be convenient to use some notations. Given any sequence
$\bm x = (x_1, x_2, \ldots)$, we denote by $\Delta \bm x$ and
$\Sigma \bm x$ the following sequences:
\begin{align*}
  (\Delta \bm x)_i \defeq & x_i - x_{i-1} & (\Sigma \bm x)_i \defeq \sum_{j=1,i}x_j
\end{align*}
where $x_0 \defeq 0$.  If $\bm x$ is a probability distribution,
meaning that $\sum_i x_i=1$, then the sequence $\bm x$ is the {\em
  Probability Density Function} (PDF), and $\Sigma \bm x$ is the {\em
  Cumulative Density Function} (CDF). With some abuse, we use the
terms PDF and CDF even when $\bm x$ is not a probability distribution.
Obviously, $\Sigma \Delta \bm x = \Delta \Sigma \bm x = \bm x$.

The following {\em summation-by-parts} holds, for any two sequences
$\bm x, \bm y$.
\begin{align*}
  \sum_{i=1,n} (\Delta \bm x)_i y_i = & x_n y_n - \sum_{i=1,n-1}x_i (\Delta \bm y)_{i+1}
\end{align*}
Using this formula, we prove:
\begin{lmm}
  Let $\bm a, \bm b$ be two, non-negative sequences, and assume that
  $\bm b$ is non-decreasing (meaning, $b_1 \geq b_2 \geq \cdots$). Let
  $\bm A \defeq \Sigma \bm a$, and let $\bm A'$ be such that
  $\bm A \leq \bm A'$.  Then, the following holds, where
  $\bm a' \defeq \Delta \bm A'$:
  \begin{align*}
    \sum_{i=1,n} a_i b_i \leq & \sum_{i=1,n} a_i' b_i
  \end{align*}
\end{lmm}

The proof follows from:
\begin{align*}
  \sum_{i=1,n} a_i b_i = & \sum_{i=1,n} (\Delta \bm A)_i b_i =  A_nb_n -\sum_{i=1,n-1}A_i(\Delta \bm b)_{i+1}\\
\leq & A_n'b_n -\sum_{i=1,n-1}A_i'(\Delta \bm b)_{i+1}
\end{align*}
which holds because $b_n \geq 0$ and $(\Delta \bm b)_i \leq 0$.  By
applying the lemma a second time, we infer that the output of a join
query~\eqref{eq:dsb:join} can be upper bounded by $\sum_i a'_i b'_i$,
where $\bm a', \bm b'$ are the PDFs of $\bm A'\geq \bm A$ and
$\bm B' \geq \bm B$.

\end{document}